\documentclass[aps,pre,reprint, superscriptaddress]{revtex4-1} 
\usepackage{amssymb}
\usepackage{amsmath} 
\usepackage{dcolumn}   
\usepackage{bm}        
\usepackage{graphicx}
\usepackage{natbib}
\usepackage{appendix}
\long\def\/*#1*/{}

\begin{document}
\title{Crystal growth from a supersaturated melt: relaxation of the solid-liquid dynamic stiffness}
\author{Francesco Turci}
\author{Tanja Schilling}
\affiliation{Theory of Soft Condensed Matter, Physics and Materials Science Research Unit, Universit\'e du Luxembourg, L-1511 Luxembourg, Luxembourg}

\date{\today}
\begin{abstract}
We discuss the growth process of a crystalline phase out of a metastable over-compressed liquid that is brought into contact with a crystalline substrate. The process is modeled by means of molecular dynamics. The particles interact via the Lennard-Jones potential and their motion is locally thermalized by Langevin dynamics. We characterize the relaxation process of the solid-liquid interface, showing that the growth speed is maximal for liquid densities above the solid coexistence density, and that the structural properties of the interface rapidly converge to equilibrium-like properties. In particular, we show that the off-equilibrium dynamic stiffness can be extracted using capillary wave theory arguments, even if the growth front moves fast compared to the typical diffusion time of the compressed liquid, and that the dynamic stiffness converges to the equilibrium stiffness in times much shorter than the diffusion time. 
\end{abstract} 

\maketitle

\section{Introduction}

When two phases of a material enter into contact, they form an interface characterized by a typical free energy cost. The \textit{interface free energy}, \textit{surface tension} or \textit{interfacial tension} is a global measure of such a cost. It encodes information on the shape, the local curvature, and the roughness of the region that distinguishes the two phases. Strictly speaking, an interfacial tension is properly defined only if the two phases are coexisting and at thermodynamic equilibrium: in this case we can define a canonical ensemble of equilibrium interfaces from which an interfacial free energy contribution can be computed.

On the theoretical side, much work has been done on the equilibrium properties of interfaces. While liquid-vapor systems have been largely discussed and understood \citep{evans1979nature,nijmeijer1988molecular,sides1999capillary,kuo2004ab,wertheim2008correlations}, liquid-liquid and solid-liquid interfaces are still under investigation. For equilibrium solid-liquid interfaces, even the very simplified cases of hard-spheres or Lennard-Jones particles are still subject of extensive studies aimed at the correct evaluation of equilibrium properties in connection with coarse-grained theories \citep{Davidchack:2006il,2012PhRvL.108v6101H,2012PhRvE..86b1404O,Wang:2013eo}.

Extending the concept of equilibrium interfacial tension to non-equilibrium is not straightforward. In fact, while it is always possible to define an interface between two distinct phases even during non-equilibrium  dynamics (both steady states and relaxation processes), it is not in general clear under which conditions the interfacial structural properties are actually related to a free energy cost or rather need to be seen from the purely kinetic point of view. Formally, it is always possible to define an \textit{effective interfacial tension} \citep{PhysRevLett.67.2013, ma1993dynamical}, for example starting from the anisotropy of the pressure tensor, from the composition gradient across the interface or, as we shall see in the following, from the fluctuations of the shape of the interface itself. There is simulation and experimental evidence that the description of growth phenomena requires the introduction of an effective \textit{dynamic tension} in order to properly stabilize the interface during the transient states for fluid-fluid interfaces \citep{2014PhRvL.112l8303T,cicuta2001capillary,PhysRevE.78.016306}, but there still is a substantial lack of consensus on the determination of the proper relaxation times related to it and their interpretation \citep{PhysRevE.86.015301,shikhmurzaev2005capillary,Henderson:2011em,Lukyanov:2013br}.

In the present study we concentrate our attention on solid-liquid interfaces and the relaxation process associated with crystallization. As compared to relaxing liquid-liquid interfaces, crystal growth poses an additional challenge due to the elastic and viscous properties of the growing, off-equilibrium solid phase. While scaling theory can be useful in order to provide general arguments about kinetic models underlying solid-on-solid growth \citep{lai1991kinetic, barabasi1995fractal,gilmer2003simulation,Yoon:1999vx,Pal:2003hk}, we devote our study to the analysis of the capillary effects related to the growth processes and their counterparts at equilibrium, in order to characterize the crossover from the dynamic to the equilibrium interfacial tension for solid-liquid interfaces.

Our study is organized as follows: in section \ref{cwtsec} we recall some fundamental concepts related to capillary wave theory (CWT) at equilibrium and out of equilibrium; in section \ref{growthsec} we present our model for heterogeneous crystal growth and provide an illustration of the growth kinetics; in section \ref{noneqsec} we proceed to the description of the dynamic interfacial properties, presenting two examples of relaxation and two independent means of numerically deriving estimates for the dynamic stiffness based on CWT; we conclude in section \ref{conclude} with the discussion of the results in comparison with the equilibrium features of the solid-liquid interface.

\section{Capillary wave theory and stiffnesses}
\label{cwtsec}
For liquid-liquid interfaces in thermodynamic equilibrium, shape flucutations are described by capillary wave theory (CWT)\citep{Buff:1965tt, evans1979nature,Fisher:1982wd,Bedeaux:1985dk,binder2001intrinsic}. CWT is derived by applying the equipartition theorem to the distribution of Fourier modes of the interfacial height fluctuations \citep{Buff:1965tt, evans1979nature,Fisher:1982wd,Bedeaux:1985dk,binder2001intrinsic}. Albeit thought and developed for gas-liquid or liquid-liquid interfaces, this simple theory has also been successfully applied to the analysis of solid-liquid interfaces  \citep{Morris:2002fk,Morris:2003ud,Rozas:2011hw}.

The main prediction of the theory, directly related to equipartition of the thermal energy over the possible degrees of freedom, is the scaling of the height-height correlation function $h(\vec{r})$ with the (two dimensional) radial position $\vec{r}=x \hat{e}_{x}+y\hat{e}_{y}$. This correlation function describes the local deviations of the interfacial height from the mean interface position. In general, for crystal-liquid interfaces, the structure of the crystal is reflected in an anisotropic form of the correlation function. Yet, for crystal orientations with $x,y$ symmetry, in the limit of short wave-vectors $\vec{q}$ one can write the power spectrum of height fluctuations as:
\begin{equation}
\langle |h_{\rm eq}(q)|^2\rangle=\frac{k_{B}T}{L^{2}\tilde{\gamma}q^{2}}
\label{cwt1}
\end{equation}
where $q=|\vec{q}|$, $\tilde{\gamma}$ is the stiffness of the interface, $T$ is the temperature, $k_B$ is Boltzmann's constant, $L$ is the lateral size of a system of square cross section $L \times L$ and $\langle\cdot\rangle$ denotes the equilibrium ensemble average. It is therefore sufficient to extract the power spectrum of the height fluctuations in order to estimate, through a best fit at low wave-vectors, the equilibrium stiffness of the interface.

Note that the stiffness, which is orientation dependent, is related to the interfacial tension by the following tensorial relation:
\begin{equation}
\tilde{\gamma}_{xy}(\textbf{n}_{z})=\gamma(\textbf{n}_{z})+\frac{\partial^{2}\gamma(\textbf{n})}{\partial\textbf{n}_x\partial\textbf{n}_y}\big|_{\textbf{n}=\textbf{n}_{z}}
\label{eq:stifftens}
\end{equation}
where $\textbf{n}_{x},\textbf{n}_{y}, \textbf{n}_{z}$ correspond to the (3d) principal axes of $\partial^{2}\gamma(\textbf{n})/\partial\textbf{n}_{i}\partial\textbf{n}_{j}|_{\textbf{n}=\textbf{n}_{z}}$.
 Such a relation allows to compare stiffnesses and tensions, once several crystal orientations are taken into account \cite{1983PhRvL..50.1077F,2012PhRvL.108v6101H}. 

One can define the temporal dynamical correlation function expressed in terms of the spatial height-height correlation function as 
\begin{equation}
g_h (\Delta t)=\langle h'(\vec{q},t) h'(\vec{q},t+\Delta t) \rangle
\label{eqgt}
\end{equation}
where $h'(\vec{q},t)=h(\vec{q},t)-\langle h(\vec{q},t)\rangle_t$ is the deviation from the time-averaged profile. This two-time correlation function is characterized by a typical decorrelation \textit{capillary time} $\tau_{\rm decorr}$, that is related to the relaxation of the interface close to equilibrium and that determines the dynamics of the modes on the interface. 

Integrating the spectrum in eq.~\ref{cwt1} between a minimum cutoff length $l_{\rm cut}$ and the linear size of the cross section $L$, and applying a convolution approximation\citep{Binder:2000vf}, one finds the scaling of the mean interfacial width:

\begin{equation}
\label{logscal}
w^{2}=w_{0}^{2}+\frac{k_{B}T}{4\tilde{\gamma}}\ln(L/l_{\rm cut}) 
\end{equation}
where $w_0$ refers to an hypothetical \textit{intrinsic} interfacial width, defined beyond any roughening due to capillary effects \citep{Buff:1965tt}. While the intrinsic interfacial width is hardly measurable, this expression allows for the computation of the interfacial stiffness $\tilde{\gamma}$ by means of finite size scaling analysis.

CWT is an appropriate theory for rough liquid-gas and liquid-liquid interfaces: in the case of liquid-solid interfaces, the applicability of the same theory is not as clear, since it makes the strong approximation that a solid, crystalline phase can be considered as an extremely viscous fluid phase, with consistent stiffnesses and capillary times \citep{HernandezGuzman:2009ia}. Yet, it has been numerically shown \citep{ZykovaTiman:2009iq} that the logarithmic scaling of CWT can be effectively used as a means to correctly estimate the interfacial stiffness for solid-liquid mixtures \citep{Hoyt:2001er,Davidchack:2006il}. 
\newline 

When two phases are put into contact,
a relaxation process of the interfacial properties is initiated, which eventually converge into the properties of the equilibrium interface as it is described by capillary wave theory.  
The shape of the interface during the relaxation process is partly determined by the nature of the driving force that creates the non-equilibrium state.
 The specific form of the power spectrum of the height-height fluctuations obeys different scaling laws in the case of sharp interfaces, linear gradients of concentration or temperature, and interfaces moving at constant velocities \citep{1993PhRvE..48.3441K,Vailati:1998tp,2000PhRvE..62.4920C}. In ref.~\cite{1993PhRvE..48.3441K} the spectrum of non-equilibrium stationary height fluctuations is derived for an interface that is moving at constant velocity $v$. It converges in the limit of small velocities $v\rightarrow 0$, to 
\begin{equation}
\langle |h_{\rm noneq}(q)|^{2}\rangle_{\rm trajectories}=\frac{k_{B}T}{L^{2}\tilde{\gamma}(q^{2}+4\pi^{2}/l^{2})}
\label{eq:specnoneq}
\end{equation}
where $l$ is a crossover length due to the chemical potential gradient and the latent heat of the liquid-to-solid transition, that has a pinning effect on fluctuations occurring on scales larger than $l$. The consequence of the driving is thus, in this limit, similar to the effect of gravity on liquid-liquid equilibrium interfaces.

As suggested by equation \ref{eq:specnoneq}, in the case of a stationary state, a non equilibrium dynamic surface tension can be derived directly via a fit of the spectrum of height fluctuations. We use this procedure in the case of a relaxation process driven by the chemical potential difference; we follow the time dependence of the spectrum and extract from it an estimate for the effective dynamic tension. This is an alternative to the computation of gradients and anisotropies across the interface, which also define dynamic stiffnesses, such as in the case of composition gradients in Kortweg's theory of effective surface tension \citep{kurtweg1901}, the pressure tensor anisotropy \citep{Lukyanov:2013br} or non-equilibrium linear response to excitations \citep{Loglio1991335,Franses1996296}.

In the following, we will analyze different relaxation dynamics related to the crystallization of a Lennard-Jones liquid in contact with a crystalline substrate using  CWT and fluctuating hydrodynamics in order to extract the dynamic behavior of the crystal-liquid interfacial stiffness that is related to the tension by eq.~\ref{eq:stifftens}.

\section{Growth Process}
\label{growthsec}
\subsection{A model for crystal growth on a substrate}
\label{modelsec}

We consider a simple model of out-of-equilibrium dynamics modeling crystal growth and the motion of a solid-liquid interface and its capillary mode fluctuations.
 We perform Molecular Dynamics simulations of crystallization of an over-compressed liquid that is in contact with a substrate, both composed of identical particles of mass $m$ interacting via a truncated and shifted Lennard-Jones potential
$u(r)=u_{\rm LJ}(r)-u_{\rm LJ}(r_{\rm cut})$ for $r<r_{\rm cut}$ and 0 otherwise, with $u_{\rm LJ}(r)=4\epsilon[(\sigma/r)^{12}-(\sigma/r)^6 ]$ and $r_{cut}=4\sigma$.

We first prepare separately an over-compressed, metastable liquid at high density and an immobile defect-free crystal of particles exposing the [100] orientation to the liquid perpendicular to the $z$ axis of the simulation box, with unit cell length $a=\sqrt[3]{4/\rho_{\rm sol}}$ where $\rho_{\rm sol}(T)$ is the coexistence density for the solid phase at a given temperature (see Appendix \ref{details} for the simulation details). We consider liquid densities $\rho$  ranging from about $0.9\rho_{\rm sol}(T)$ up to maximum $1.4\rho_{\rm sol}(T)$ avoiding configurations where crystallites are present due to homogeneous nucleation. 

The system undergoes isothermal and isochoric Langevin dynamics -- isochoric, to avoid the additional complication due to effects of a barostat on the dynamics. This implies that the density of the over compressed liquid does not remain constant during the growth of the crystal: the interface progresses along the z-axis driven by a time-dependent chemical potential difference that gradually diminishes, eventually leading to a condition for which the liquid reaches the equilibrium coexistence density. 

\subsection{Growth Profiles and Growth Speeds}
\label{profilesec}
\begin{figure}[t]
\centering
\includegraphics[width=\columnwidth]{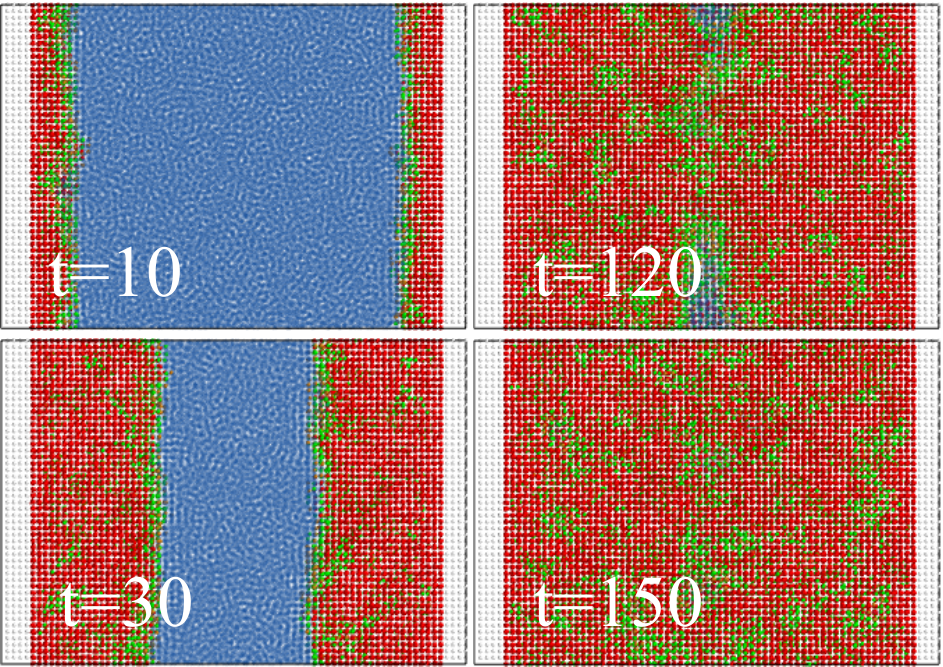}
\caption{(color online) Four snapshots representing the time-evolution of the crystal growth on top of a perfectly ordered template (gray particles) from a lateral side point of view. The density of the liquid is $\rho/\rho_{\rm sol}=1$, the temperature is $T=0.7793\epsilon$ and the system has dimensions $25\times25\times64 a^{3}$, with time measured in units of $t_{\rm LJ}:=\sqrt{m\sigma^{2}/\epsilon}$. The degree of crystallinity, determined using the $\bar{q}_{4},\bar{q}_{6}$ order parameters, blends from essentially fcc-like (red) to liquid (blue) through a 2-3 layers thick interface of hcp-like particles.}
\label{snaps}
\end{figure}

In order to distinguish the crystal from the liquid phase we use a set of observables that measure the degree of local order in the neighborhood of every particle. We use the so called locally averaged bond order parameters $\bar{q}_{4},\bar{q}_{6}$, inspired by Steinhardt's order parameters \citep{1983PhRvB..28..784S} and defined in \citep{Lechner:2008vz} (see the Appendix for more details). These order parameters take into account the first and the second shell of neighbours around a given particle: this allows for the distinction between different crystalline structures (face centered cubic, body centered cubic, hexagonal close packed). 
Using these observables, we follow the phase transition from the over-compressed liquid, tracking the gradual increase of crystalline order. As shown in fig.~\ref{snaps}, the growth of the crystal involves the formation of a highly ordered region of fcc structures separated from the remaining liquid particles by few layers of hcp-like particles. However, the growing crystal relaxes only slowly towards its equilibrium configuration. At any time, there are extended clusters of impurities (green regions), a signature of the fact that the relaxation inside the crystal proceeds much more slowly than the formation of new crystalline layers. 
 
\begin{figure}[t]
\centering
\includegraphics[width=\columnwidth]{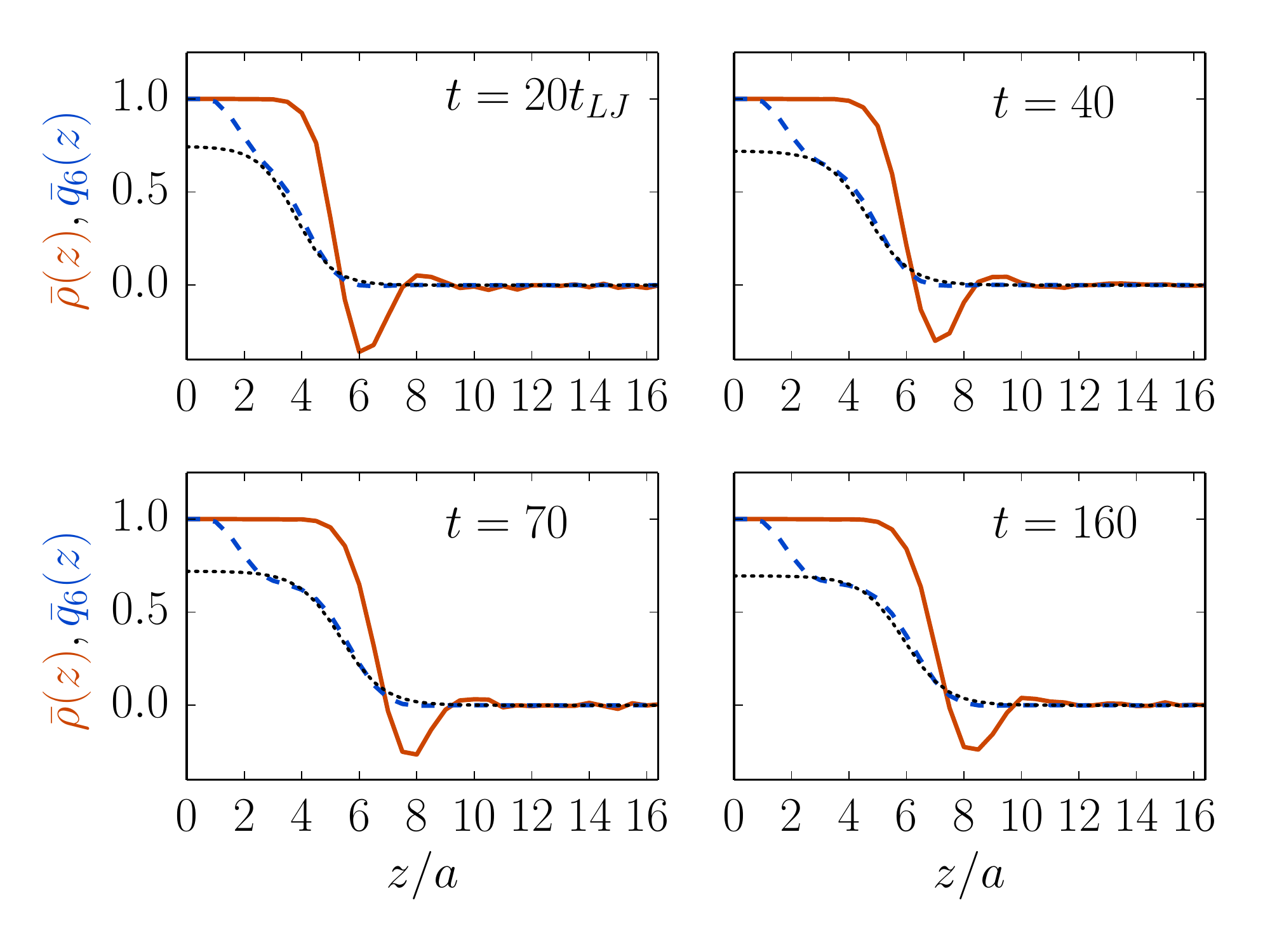}
\caption{(color online) Time evolution of the coarse-grained laterally averaged density (red continuous line) and $\bar{q}_6$ (dashed blue) profiles, normalized in such a way that the perfect crystal phase has value 1 and the bulk liquid takes value 0. The black dotted line corresponds to the fit of the $\bar{q}_6$ profiles according to eq.~\ref{eq:tanh} in order to determine the position of the liquid-solid interface and its width. Notice that the $\bar{q}_6$ profile shows two interfaces (wall-crystal and crystal-liquid respectively): we concentrate our study on the properties of the moving crystal-liquid interface.}
\label{profs}
\end{figure}

The local order observables permit to identify the interface and describe its dynamics. It is sufficient to track the coarse-grained laterally averaged $\langle\bar{q}_{6}(z,t)\rangle_{x,y}$ profile in order to determine the position and the width of the interface (we denote with $\langle\cdot\rangle_{x,y}$ the average over lateral cross sections of the system). The profile is modeled with a hyperbolic tangent function
\begin{equation}
\langle\bar{q}_{6}(z,t)\rangle_{x,y}=C_{1}+C_{2}\tanh\left(\frac{z-z_{0}(t)}{w(t)}\right)
\label{eq:tanh}
\end{equation}
which provides the values of the position $z_{0}(t)$ and the width $w(t)$ via a four parameters best fit procedure (see fig.~\ref{profs}).

In order to improve the statistics, we consider a sample of 100 independent crystallization trajectories, where we restart the simulation from different equal-energy configurations of the metastable liquid. Density profiles, local order parameter profiles, interface trajectories and alike quantities result from a same-time average over such a sample, with time measured in Lennard-Jones units.

\begin{figure}
\centering
\includegraphics[scale=1]{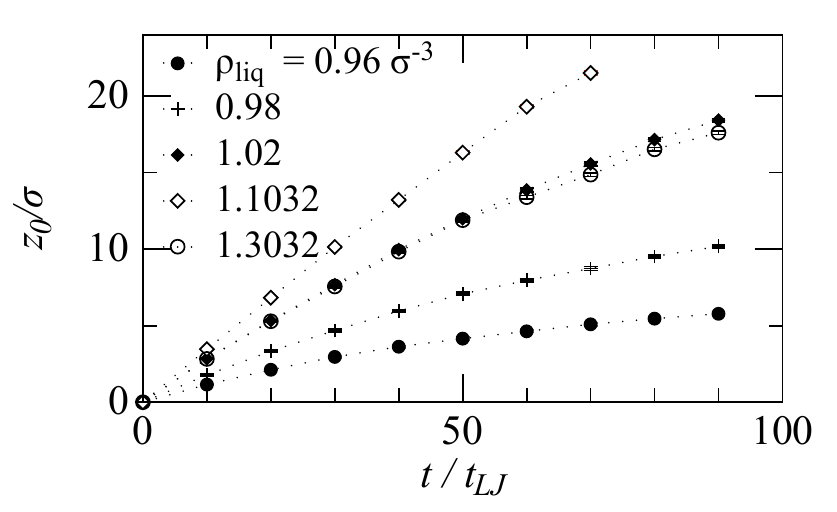}
\caption{Average interface position (extracted from best fits of eq.~\ref{eq:tanh} to the $\bar{q}_6$ profiles) for a system at temperature $T=1.061\epsilon$ and solid coexistence density $\rho_{\rm sol}=1.0174\sigma^{-3}$ for different initial liquid densities $\rho_{\rm liq}$. Note that the speed of the interface is non-monotonic in $T$, i.e. the curve that corresponds to $T=1.3032\epsilon$ lies below the  $T=1.1032\epsilon$ curve.}
\label{gcurv}
\end{figure}

We use the crystallinity profiles defined by eq.~\ref{eq:tanh} in order to track the position and the speed of the growing crystal front.
In fig.~\ref{gcurv} we show average interface trajectories $z_{0}(t)$.  Depending on the initial difference in densities between the substrate and the over-compressed liquid, saturation of the interface position to its final value is reached at different times after an initial linear growth regime. We are interested in the characteristic initial velocity $v_0$, when the system is far from equilibrium and driven by a large chemical potential difference.
For the largest over-compressions the initial linear growth regime lasts sufficiently long that the crystal completely fills the box.

The crystallinity profiles shown in fig.~\ref{profs} can be compared with the laterally averaged density profiles $\bar{\rho}(z)=\langle\rho(x,y,z)\rangle_{x,y}$. While the local bond order parameter monotonically connects the solid to the liquid region, the averaged density profile is marked by the presence of a depletion zone in the vicinity of the interface (similarly to what has been observed in equilibrium for both hard spheres \citep{Sandomirski:2011tf} and Lennard-Jones particles \citep{Huitema:1999ka}), which is partly compensated by a higher density of the first crystal layers immediately in contact with the liquid. The depth of the depletion zone depends on time, while the local bond order profile simply translates along the z axis.
The change in density spatially precedes the change in $\bar{q}_6$, effectively determining two interface positions that differ by approximately one unit cell.

\begin{figure}[h!tbp]
\centering
\includegraphics[scale=1]{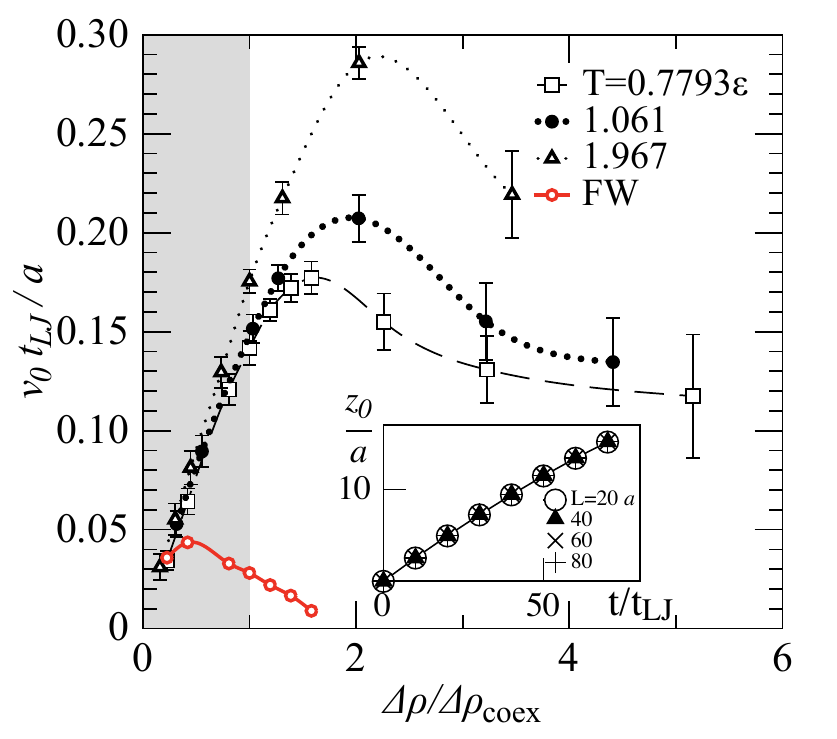}
\caption{Velocity of the crystal-liquid interface as a function of the ratio $\Delta\rho/\Delta\rho_{\rm coex}=(\rho-\rho_{\rm liq})/(\rho_{\rm sol}-\rho_{\rm liq})$ for different densities $\rho$ of the metastable, over-compressed liquid. The gray area corresponds to the range of coexistence densities, above which the crystalline phase is the only stable phase. Inset: evolution of the position of the interface as a function of time for $\rho/\rho_{\rm sol}=1, T=0.7793\epsilon$ for different linear lengths $L_{x}=L_{y}=20a,40a,60a,80a$: finite size effects play only a marginal role in the determination of the growth velocity.}
\label{speeds}
\end{figure}

By fitting the profiles we obtain the initial growth velocity as a function of the initial liquid density (see fig.~\ref{speeds}). The initial growth velocity has a maximum at liquid densities slightly higher than the coexistence solid density $\rho_{\rm sol}$, in contrast to what has been reported in the case of hard-spheres \citep{Sandomirski:2011tf}. For both temperatures considered here, the maximum is reached at liquid densities such that $\Delta\rho/\Delta\rho_{\rm coex}\approx 2$, with $\Delta\rho/\Delta\rho_{\rm coex}=(\rho-\rho_{\rm liq})/(\rho_{\rm sol}-\rho_{\rm liq})$, where $\rho_{\rm liq},\rho_{\rm sol}$ indicate the equilibrium coexistence densities for the liquid and the solid phase respectively. At even higher densities the competition between the increasing driving force due to the chemical potential difference and the dramatic decrease of the diffusion constant \citep{Meier:2004fp} leads to a decrease of the growth velocity. Yet, compared to the hard-spheres case, the decrease is much less pronounced.  Notice that the growth velocity is generally very fast compared to the typical diffusion time: for example, the peak velocity for $T=0.7793\epsilon$ is $v_{\rm peak}\approx 0.17 a/t_{LJ}=0.11\sigma/t_{LJ}$, to be compared with the diffusion time $\tau_D=\sigma^2/D\approx 50 t_{\rm LJ}$ and velocity $v_{D}=\sigma/\tau_{D}\approx 0.02\sigma/t_{LJ}$. (As the homogeneous nucleation rate rapidly increases with the degree of over-compression, we end our analysis at a density for which the velocity is still about $60\%$ of the peak velocity.)

A simple model that reproduces some qualitative features of the growth velocities is the classical Wilson-Frenkel law \citep{Jackson:2002kc, wilson1900,frenkel1932}. It results from an estimated balance between the rate of activated attachment and detachment of the liquid particles on the crystal surface
\begin{equation}
v(T)=A_{\rm kin}\left[ 1- \exp\left(-\frac{g_{\rm sol}-g_{\rm liq}}{k_B T}\right)\right]
\label{eq:FrenkelWilson}
\end{equation}
where $A_{\rm kin}$ is a kinetic prefactor proportional to the diffusion coefficient $D$, and $g_{\rm sol}, g_{\rm liq}$ are the solid and liquid free-energies respectively. Using accurate simulation data for the Lennard-Jones equations of state and coexistence densities \citep{2007JChPh.127j4504M}, we plug into eq.~\ref{eq:FrenkelWilson} the corresponding values for the free energy differences as a function of the liquid density and, using the diffusion coefficient computed by Molecular Dynamics simulation of large, liquid bulk systems, we obtain the red dots of figure \ref{speeds} (rescaled such that the lowest density point coincides with the velocity curves from simulation). We observe that the Wilson-Frenkel law does not allow to reproduce the large initial increase in growth speed. In fact, the resulting estimate for the speed is largely dominated by the decrease of the diffusion coefficient at large densities, which is not compensated by the chemical potential difference term. 

The failure of the Wilson-Frenkel approach is not surprising, since it is based on bulk, close to equilibrium properties, and it has already been demonstrated \citep{Kerrache:2008jr} that the specific, non-diffusive dynamics at the interface may lead to understimations of the actual velocity of at least one order of magnitude. More than that, one should also consider that the equilibrium free energies can only be a rough estimate of the actual driving force acting between the two phases of the present model, where some relaxation mechanisms occur on time scales far beyond the actual duration of the simulation. Finally, the formation of the initial layers is very fast, with barely no nucleation time related to it, thus the activation barrier seems to be suppressed \citep{Turci:2014kd}.

\begin{figure*}[t!]
\includegraphics[width=\textwidth]{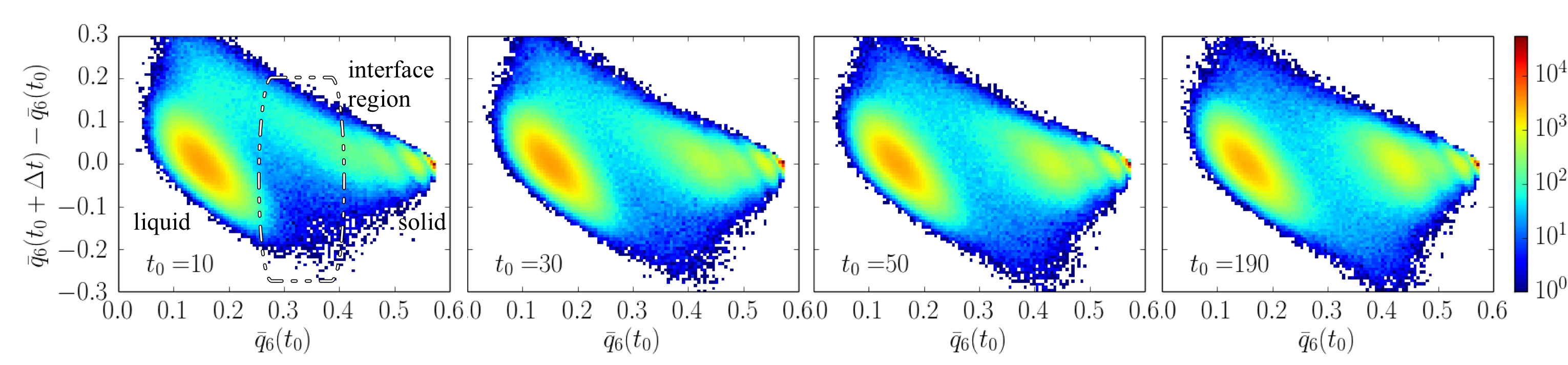}
\caption{(color online) Time evolution of the joint probability distribution $P(\bar{q}_6, \Delta \bar{q}_6)$ for a system of linear size $L=80a$ at density $\rho/\rho_{\rm sol}=0.96$ and temperature $T=1.967\epsilon$, computed for a time-interval of $\Delta t=\tau_D/5$.  }
\label{qsdiff}
\end{figure*}

\section{Non-equilibrium Interfacial Stiffness}
\label{noneqsec}
\subsection{Small chemical potential difference}
If the chemical potential difference between the liquid and the crystal is small from the beginning, the growing crystal rapidly depletes the liquid, and the motion of the interface slows down quickly. Under these conditions we can observe the relaxation of the interfacial properties into their equilbrium values on the scale of our simulation.

We set $T=1.967\epsilon$, the initial liquid density $\rho_0=1.08\sigma^{-3}$, and $\rho_{\rm sol}=1.1285\sigma^{-3}$. Using the equation of state fitted in \citep{2007JChPh.127j4504M} we estimate the initial chemical potential difference between the solid and the liquid phase to be $\mu_{\rm sol}-\mu_0=-1.99 \epsilon$ per particle. Under these conditions, growth stops after approximately 16 crystalline layers are formed.
We simulated systems of squared cross section $L\times L$ with $L=25,30,35,40,45,50,60,80\; a$ and longitudinal length $L_{z}$ such that we always have 64 fcc layers between the two initial interfaces. We extracted the time evolution of $w(t)$ (eqn.~\ref{eq:tanh}) averaged over 100 trajectories, each of which represents the growth of two opposite crystal fronts that we assume to be independent (which is true if the liquid region between the interfaces remains much larger than the typical width of the interface and the correlation length in the liquid). Regardless of the area of the cross section, after a rapid initial increase, the interfacial width saturates.

The first remarkable feature is that the whole process is extremely fast: if we call $\tau_D=\sigma^2/D$ the self-diffusion time of a particle in the over-compressed liquid, the interface attains its final position $z_0$ in a time of about $2-3\tau_D$. Thus the growth process is not dominated by the diffusion of liquid material in the vicinity of the solid-liquid interface, but by an adsorption mechanism. 

Before we analyse the evolution of the interfacial width, we briefly discuss the adsoprtion mechanism. Fig.~\ref{qsdiff} shows the joint probability distribution $P( \bar{q}_6,\Delta \bar{q}_6)$ for the difference $\Delta q_6=\bar{q}_6(t_0+\Delta t)-\bar{q}_6(t_0)$ for different times $t_0$, with a time interval of $\Delta t=10 t_{\rm LJ}\approx\tau_D/5$. As expected, the liquid distribution is centered around $\bar{q}_6\approx 0.16$ while the solid region shows the fcc peak and the signatures of the growing crystal around $\bar{q}_6=0.5$ (consistently with the typical values for Lennard-Jones bulk liquids \citep{Lechner:2008vz}). The peaks of the distribution have $\Delta \bar{q}_6=0$, because the majority of the particles are either in the liquid or in the crystal and do not change phase. Particles in the bulk, that have exceptionally low/high values of $\bar{q}_6$ are likely to move closer to the average bulk value of their phase. This is the origin of the oblique tilt in the distributions. However, particles that lie close to the interface and are incorporated into the crystal show a different dynamics and are responsible for the main differences that one can observe between the four panels: At early times the distribution is strongly asymmetric, particles with $0.2<\bar{q}_6<0.4$ are extremely likely to have an increase in local order of more that $60\%$ in the given $\Delta t$, while the reverse process is suppressed (blue dip in the bottom-centre of the first panel). This asymmetry is significantly reduced already $2\Delta t$ later at $t_0=30 t_{\rm LJ}$ and there is no noticeable difference between the last two panels, showing that the local order around interface particles fluctuates in a quasi-equilibrium manner already very early. 

We conclude therefore that while at very early times the presence of the template accelerates the formation of new crystalline layers in such a way that typically liquid particles ($\bar{q}_6 \approx 0.16$) are converted into solid particles ($\Delta \bar{q}_6\approx=+0.2$) but never converted back, the later growth dynamics (after about $1\tau_D$) consists of a balanced, close to equilibrium exchange between liquid and solid particles in the vicinity of the interface.

\begin{figure}[t]
\centering
\includegraphics[scale=1]{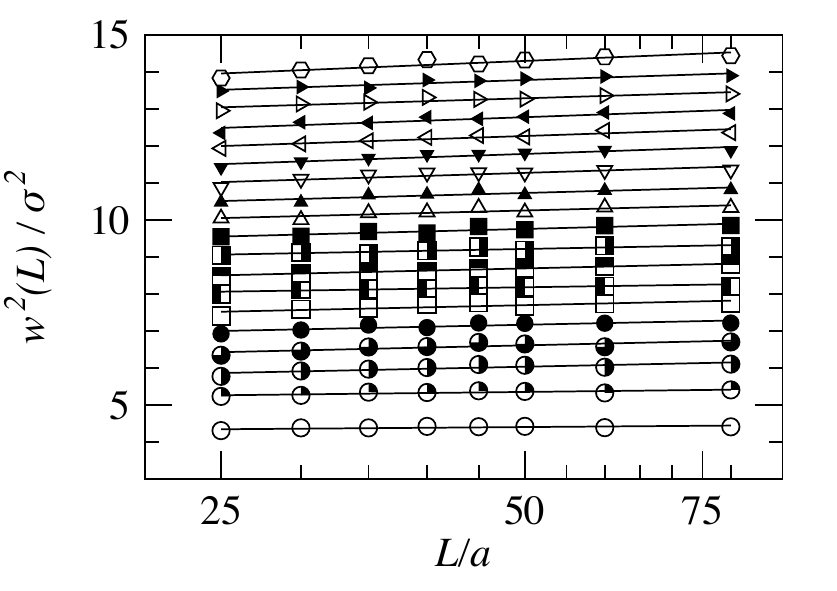}
\caption{Scaling of the the interface width with the system size (logarithmic scale on the horizontal axis) at different times going from $t=10 t_{\rm LJ}$ to $t=190 t_{\rm LJ}$, bottom to top. For ease of visualization, the curves are shifted by a constant amount proportional to $t$. }
\label{wL}
\end{figure}

\begin{figure}[bt]
\includegraphics[scale=1]{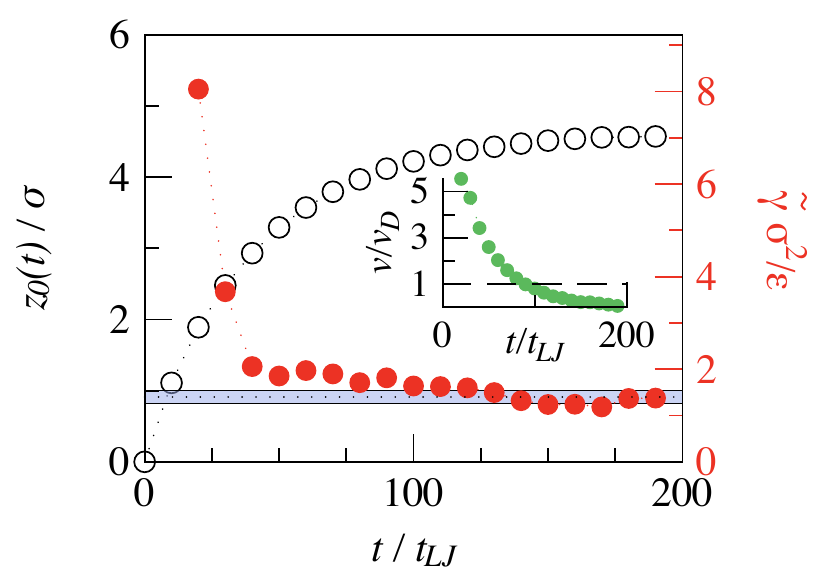}
\caption{(color online) Time evolution of the interface position (empty circles) and the interfacial stiffness (red circles) as computed from the width scaling with the lateral system size in the case of small chemical potential difference. The horizontal dotted line denotes the equilibrium value of the stiffness (within an interval of confidence of 3 standard deviations), computed from the low $q$ limit of the heigh-height correlation function. In the inset, the interface speed (green dots) compared with the diffusion speed (dashed lines) $v_D=D/\sigma$: the two cross at approximately $t\approx 100 t_{\rm LJ}$.  }
\label{z0gamma}
\end{figure}

Now we return to the analysis of the width. At equilibrium eq.~\ref{logscal} has to hold, which defines an equilibrium stiffness $\tilde{\gamma}_{eq}$. This suggests to define a \textit{dynamic stiffness} $\tilde{\gamma}(t)$ extracted from the scaling of $w_L(t)$ with the lateral linear size $L$ at a time $t$. In fig.~\ref{wL} we see, for the range of lateral sizes considered here, a logarithmic scaling of the width at any time. Based on this we compute $\tilde{\gamma}(t)$ as shown in fig.~\ref{z0gamma}. Ignoring the very early times (for which an interface width cannot be defined using eq.~\ref{eq:tanh}), we see that the dynamic stiffness starts from very high values and rapidly relaxes to an average value that we interpret as the equilibrium stiffness $\tilde{\gamma}_{eq}$.

\begin{figure}[bt]
\centering
\includegraphics[scale=1]{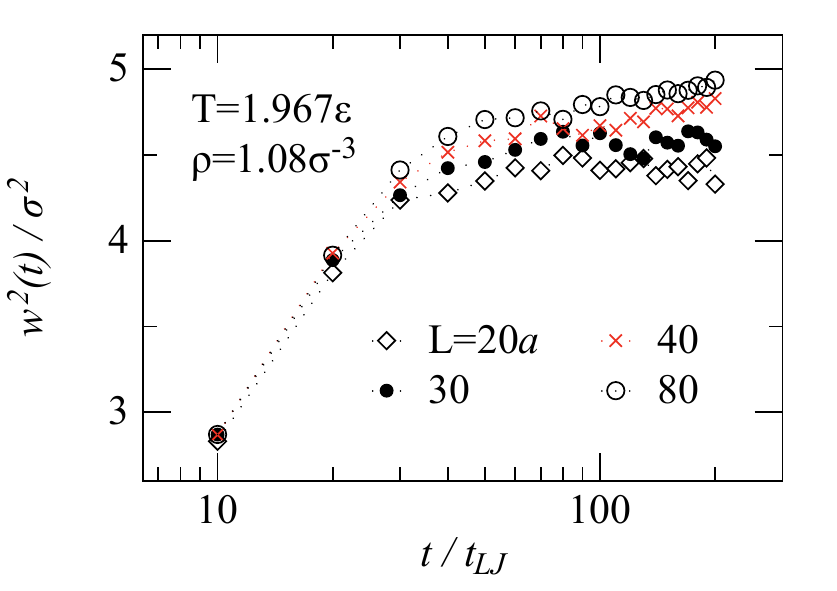}
\caption{(color online) Time evolution of the square of the interface width determined from an hyperbolic tangent fit of the $\bar{q}_{6}$ profiles as a function of time for different representative system sizes at density $\rho/\rho_{\rm sol}=0.96$ and temperature $T=1.967\epsilon$.}
\label{w(t)}
\end{figure}

\begin{figure}[tb]
\centering
\includegraphics[width=0.95\columnwidth]{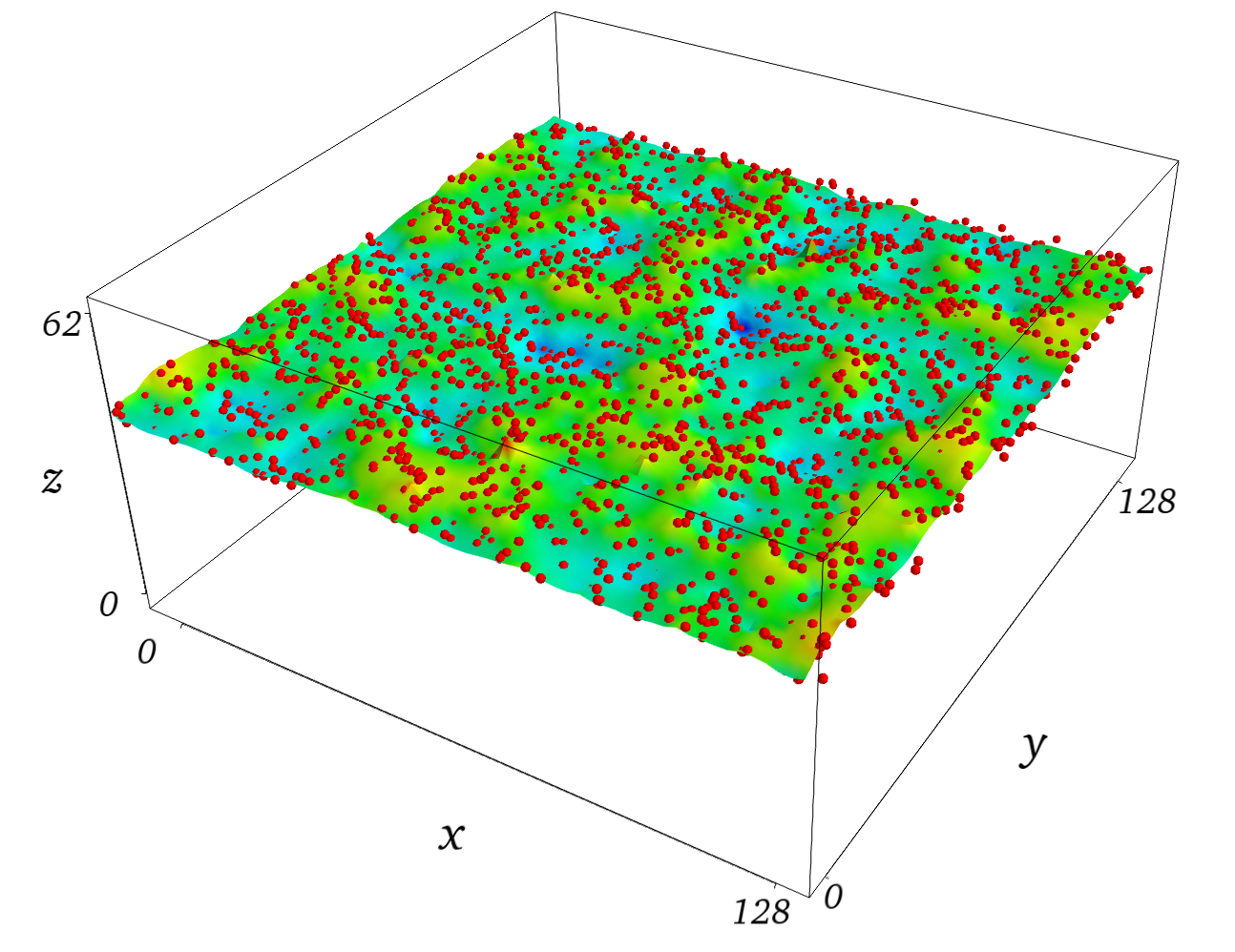}
\caption{(color online) Interface particles and interpolated interface structure during the crystal growth using the Shepard's method. On a cross section of $L_{x}\times L_{y}=6400a^{2}$ roughly 3000 particles (red dots) are isolated according to their degree of crystallinity (see text) and are the pivot for the computation of the height-height correlation function $\langle|h(\vec{q})|^{2}\rangle$. }
\label{surf}
\end{figure}
\begin{figure}[t]
\centering
\includegraphics[scale=1]{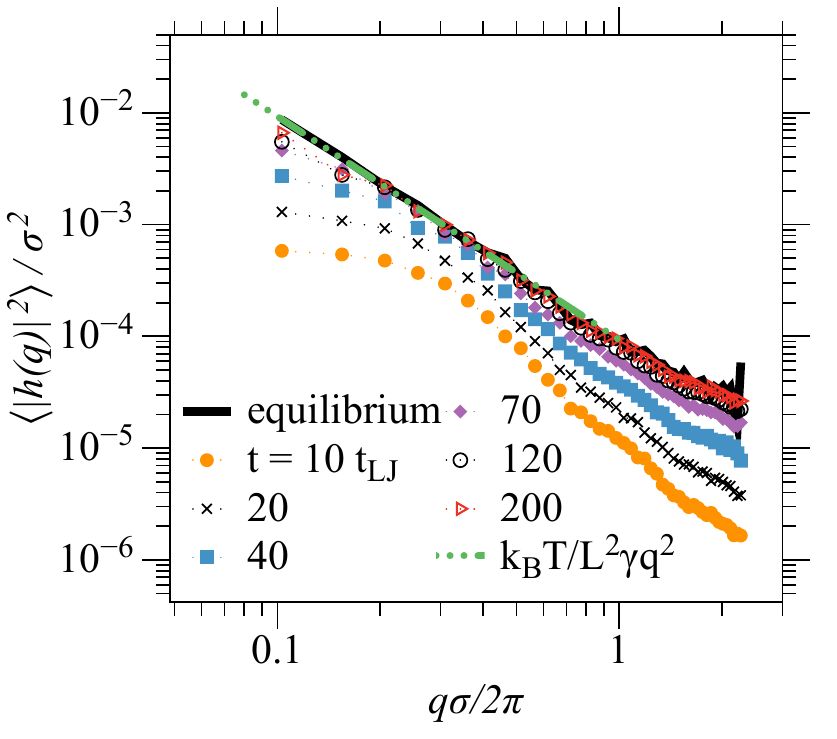}
\caption{(color online) Power spectrum of the height fluctuations for the model with small supersaturation. Notice the rapid convergence to the equilibrium profile (continuous black profile).}
\label{eqspect}
\end{figure}

To test this interpretation, we compute the stiffness from the height-height correlations.
We construct the discretized height function $h_{ij}$ by mapping interface particles onto a regular grid interpolating by Shepard's method \citep{Rozas:2011hw}. In detail, we use the local bond order parameters in order to distinguish solid and liquid particles and select only those particles that satisfy $\bar{q}_{4}>0.025$ and $\bar{q}_{6}<0.37$ and have a number of solid neighbors $5<n_{\rm sol}<12$ \citep{q6q6}. For the largest cross section $L\times L=6400a^{2}$ this procedure identifies roughly 3000 particles at the interface between the solid and the liquid (see fig.~\ref{surf}). Then, the $(x_{i},y_{i})$ coordinates of the particles are cast onto a regularly spaced grid of spacing $\Delta x=\Delta y= 2\sigma$ on top of which the $z_{j}^{\rm grid}$ coordinate is obtained as $z_{j}^{\rm grid}=\sum_{i}w_{ij}z_{i}/\sum_{i}1/r_{ij}$, where the summation runs over particles neighboring within a shell of radius $4\Delta x$ from the grid point $j$ and $r_{ij}=\sqrt{(x_{i}-x_{j})^{2}+(y_{i}-y_{j})^{2}}$. One obtains then the discretized height $h^{grid}=z^{grid}-\langle z^{grid}\rangle$ and by a Discrete Fourier Transform one derives the power spectrum $\langle |h(q)|^{2}\rangle_{\rm trajectories}$ shown in fig.~\ref{eqspect}.
\\

At early times, the spectrum shows a large plateau at low $q$ and at higher $q$ it decreases as $q^{-\alpha}$ with $\alpha\approx 2.5$, faster than the equilibrium CWT prediction. One can interpret the plateau as an effect of the pinning force due to the chemical potential difference between the liquid and the crystal phase, that increases the cost of large amplitude, long range fluctuations. The quick convergence towards a quasi-equilibrium behavior is reflected in the time evolution of the spectra: the plateau region is shortened rapidly and at intermediate $q$ the $1/q^2$ decay is established; the amplitude of the fluctuations (related to the broadening of the interface width) increases and already at $t=70 t_{\rm LJ}$ (ca.~1.5 $\tau_D$) the spectrum is hardly distinguishable from the equilibrium spectrum. Yet, the interface is still moving with a speed that is fast compared to diffusion, and new particles are added to the crystal. This appears to affect only the very low $q$ region, as a remnant of the effect of the pinning force and the initially flat profile.

Fitting the equilibrium profile with eq.~\ref{cwt1} we obtain an estimate for the equilibrium stiffness $\tilde{\gamma}_{eq}=1.42(3) \epsilon/\sigma^{2}$ that we plot as a reference value also in fig.~\ref{z0gamma}. 

\begin{figure}[t]
\centering
\includegraphics[scale=1]{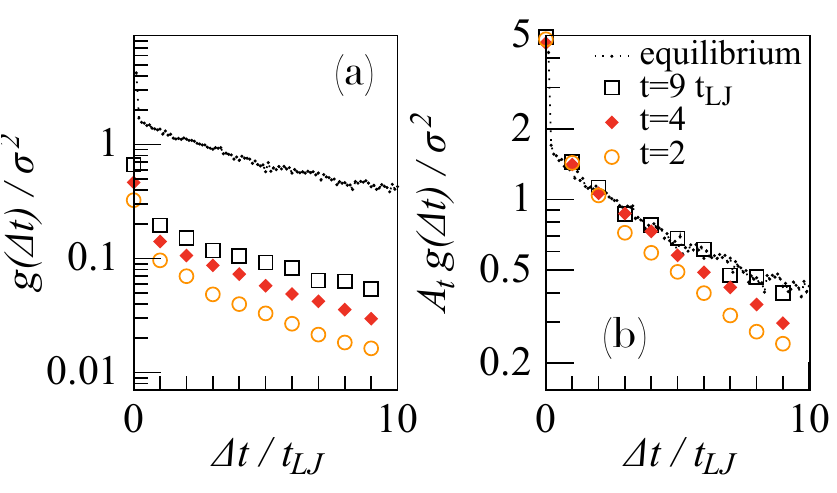}
\caption{(color online) Two-times correlation functions for the small chemical potential regime: the equilibrium curve (black dots) is compared with the time dependent ones (a). In panel (b) a scale factor $A_{t}$ is used in order to compare the trends.}
\label{g(t)}
\end{figure}

The scaling of the widths suggests that the interface rapidly evolves into a quasi-equilibrium regime. We underpin this observation by computing the typical relaxation time of the capillary fluctuations. We sample the function 
\begin{equation}
g_{t}(\Delta t)=\langle h(\vec{q},t) h(\vec{q},t+\Delta t) \rangle_{\rm trajectories}
\end{equation}
parametrized by the starting time $t$, and where the average is performed on equal-time configurations of distinct trajectories. We consider several initial spectra at different times $t$ and follow the subsequent evolution of the correlation function $g_t (\Delta t)$ in the later time steps. In fig.~\ref{g(t)} we show the decay of these correlation functions, compared to the decay of equilibrium fluctuations defined in \ref{eqgt}: in particular, in fig.~\ref{g(t)}(b) we show, by rescaling the $g_t(\Delta t)$ onto the equilibrium $g_{\rm eq}(\Delta t)$, that, while the initial relaxation process is marked by faster decays (shorter capillary times), long before the velocity of the interface decreases below the diffusion speed, the capillary time already converges to its equilibrium value.

\subsection{Large chemical potential difference}

\begin{figure}[tb]
\centering
\includegraphics[width=\columnwidth]{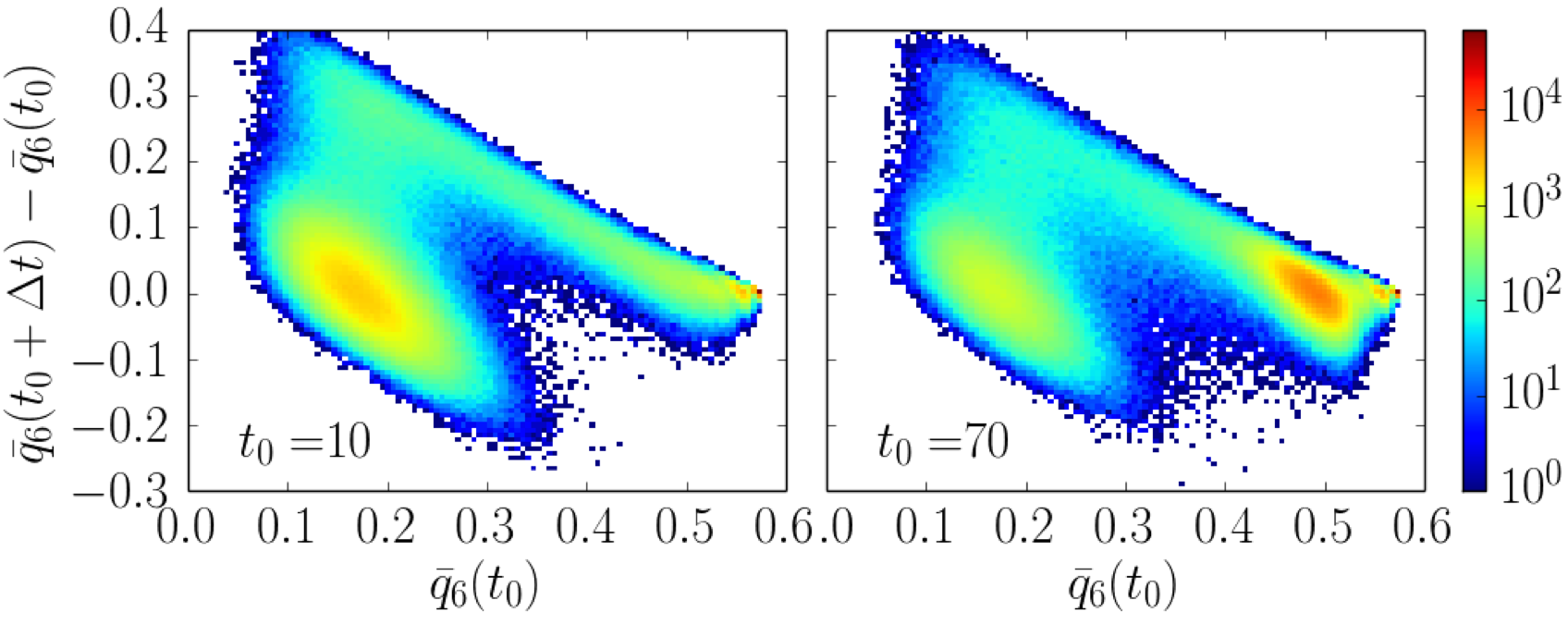}
\caption{(color online) Joint probability distribution $P(\bar{q}_6, \Delta \bar{q}_6)$ at initial times ($t_0=10 t_{\rm LJ}$) and just before the merge of the two oppositely growing interfaces ($t_0=70 t_{\rm LJ}$) for the strong driving case. Notice that the distribution is strongly asymmetric at both early and late times. Adsorportion is therefore the main mechanism, rapidly transforming liquid particles into crystalline particles, with no backward process.}
\label{figqsdeltas2}
\end{figure}

If the chemical potential difference is large, growth is arrested by the contact between the two interfaces that move in opposite directions. Up to that point they move with a large, almost steady velocity, and the growth mechanism is strong adsorption, see fig.~\ref{figqsdeltas2}. In the following we explore the evolution of the interfacial height-height correlation for the fast, initial part of the growth process. 

\begin{figure}[b]
\centering
\includegraphics[scale=1]{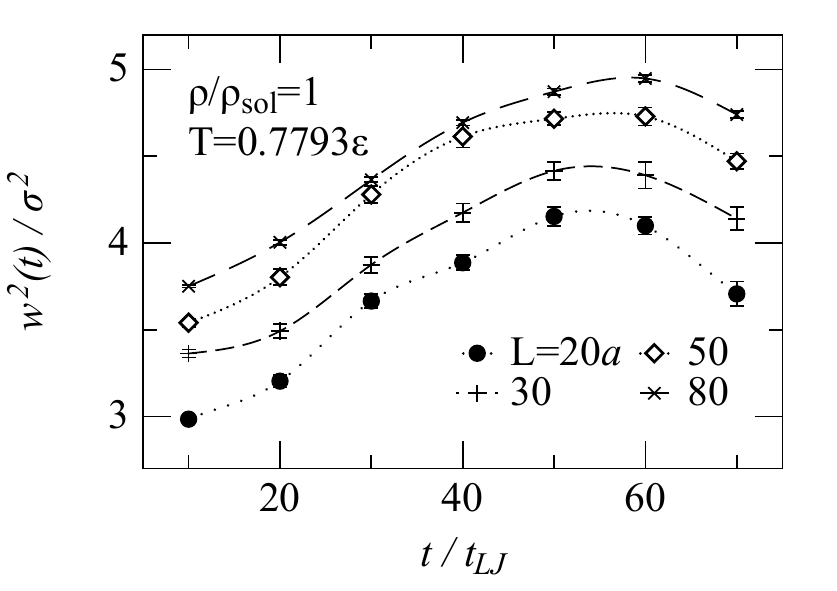}
\caption{Time evolution of the square of the interface width determined from a hyperbolic tangent fit of the $\bar{q}_{6}$ profiles as a function of time for different representative system sizes at density $\rho/\rho_{\rm sol}=1$ and temperature $T=0.7793\epsilon$ per particle. The maximum is reached when the interfaces begin to merge. The dashed lines are guides for the eye.}
\label{w(t)2}
\end{figure}

\begin{figure}[t]
\centering
\includegraphics[scale=1]{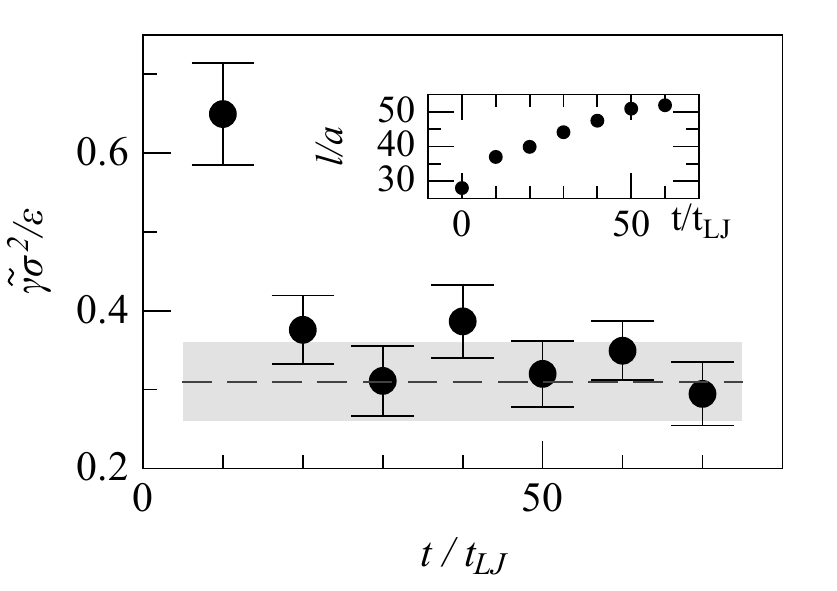}
\caption{Dynamic stiffness as a function of time for the strongly driven system. The obtained values are in agreement with equilibrium values (gray-shaded area). In the inset: crossover scale $l$ as a function of time in lattice constant units. Notice that throughout the relaxation, the crossover scale is finite and below the system cross-section size $L=80a$.}
\label{stiff}
\end{figure}

Using the same techniques as applied in the case of smaller driving forces, we track the interface widths for a system at temperature $T=0.7793\epsilon$ with $\rho_0=\rho_{\rm sol}=0.9732\sigma^{-3}$ (see fig.~\ref{w(t)2}) corresponding to a theoretical chemical potential difference $\mu_{\rm sol}-\mu_0=-3.37\epsilon$ per particle. Exploiting the system size scaling, we compute the dynamic surface stiffness (fig.~\ref{stiff}). Similarly to the previous smaller $\Delta\mu$ case, we compare the resulting stiffness with the one obtained from the capillary spectrum of an equilibrated system at the same temperature. Again, after an initial rapid drop, even at very early times the dynamic stiffness computed from the width scaling is in agreement with the equilibrium behavior. 

\begin{figure}[t]
\centering
\includegraphics[scale=1]{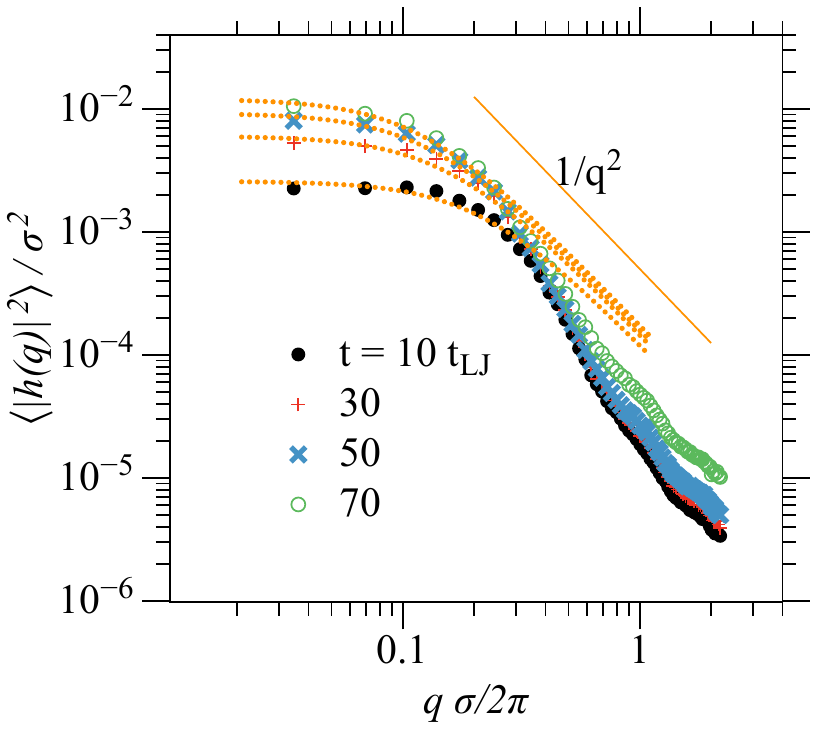}
\caption{(color online) Height-height correlation function $\langle |h(q)|^{2}\rangle$ for the growing crystal-liquid interface at $\rho/\rho_{sol}=1.$ and $T=0.7793\epsilon$. The dotted lines represent the expression in eq.~\ref{eq:specnoneq} where we use the value of the interfacial stiffness $\tilde{\gamma}$ independently obtained from the analysis of the widths of the $\bar{q}_{6}(z)$ profiles.}
\label{spectr}
\end{figure}

The signature of the non-equilibrium dynamics can be found only in the detailed shape of the height-height spectrum (fig.~\ref{spectr}): the plateau region at low $q$ is always present, due to the non-negligible chemical potential difference (corresponding to the high speed of the interface). The expected CWT scaling does not have sufficient time to emerge. Thus we test how the stationary, low-velocity limit of eq.~\ref{eq:specnoneq} applies to the measured spectra. We perform a single parameter fit using, as input, the stiffness obtained from the $L$ scaling of the widths extracted from $\bar{q}_6(z)$ profiles. We get as a unique fitting parameter the crossover length $l$ the time evolution of which (far from being stationary) is shown in the inset of fig.~\ref{stiff}. This crossover length brings the signature of the slowly reducing pinning force given by the chemical potential difference, and it diverges in the limit of thermodynamic equilibrium when $\Delta\mu=0\epsilon$.

Thus, the spectra (and the dynamic stiffness) reveal the non-stationary relaxation process of the growing crystal lattice. In particular, the fully non-equilibrium spectra are clearly different from the predictions of CWT or fluctuating hydrodynamics \citep{1993PhRvE..48.3441K}. Yet, when dealing with globally averaged quantities such as the interface width, it is still possible to observe the logarithmic scaling with the system size, which permits to extract a dynamic stiffness still consistent with the equilibrium reference value.

\section{Conclusions}
\label{conclude}
Simulating a simple model under isochoric, isothermal conditions, we identify some characteristic features of crystallization from an undercooled liquid on a substrate: First, we have observed that the fastest initial growth speed is achieved when the liquid density is even higher than the crystalline template density.  Then we have discussed two dynamical regimes at small and large driving forces, fixed by the chemical potential difference between the substrate and the over-compressed liquid: at small driving forces we see that liquid particles get initially adsorbed onto the growing crystal and, when the interface velocity relaxes and becomes of the order of the diffusion speed, a quasi-equilibrium regime is attained. A dynamic stiffness can be defined using the scaling of the interface width. And the height-height fluctuations rapidly converge to a CWT-like scaling regime. In the strongly driven case, the spectra are very different from the CWT regime at any time but it is still possible to compute a dynamic stiffness that is of the same order of magnitude of the equilibrium reference stiffness.

The possibility of treating the driven dynamics with concepts derived from CWT is in line with previous studies of simplified models for liquid-liquid interfaces out of equilibrium, such as under shear \citep{Smith:2010da}. 
The fact that even in a strongly driven dynamical regime the effective dynamic stiffness appears to rapidly converge towards the corresponding equilibrium value is particularly promising for a coarse-grained description of the dynamic that uses equilibrium properties in input, such as the Dynamic Density Functional Theory (DDFT) or, at a higher level, a Phase-field crystal approach\citep{Emmerich:2012ko}. DDFT has been successfully applied for nonequilibrium dynamics of hard spheres, for instance in the case of sedimentation \citep{2007PhRvL..98r8304R}, and the present results suggest that a proper choice of the functional form for the free energy functional could reproduce the dynamics of the growing front, including the time evolution of the dynamic tension.
 
\section*{Acknowledgements}
We thank Martin Oettel for insightful discussions. This project has been financially supported by the INTER/DFG/11/06 on Crystallization. Computer simulations presented in this paper were carried out using the HPC facility of the University of Luxembourg.

\begin{appendix}
\section{Simulation details}
\label{details}

We simulated Lennard-Jones liquid particles at temperature $T=0.7793, 1.061, 1967\epsilon$ and variable densities $\rho$ confined by two fcc walls of fixed particles. The walls had a surface of $A=30\times30 a^{2}$ where $a$ is the fcc lattice spacing $a=\sqrt[3]{4/\rho_{\rm lattice}}$, with $\rho_{\rm lattice}=\rho_{\rm sol}(T)$ corresponding to the solid coexistence density at the given temperature.  The wall particles form three crystalline layers per surface and interact with the liquid particles with the identical interaction potential holding between liquid particles, with cutoff at $r_{cut}=4\sigma$. This value is particularly appropriate for the correct sampling of crystalline structures. The distance along the $z$ axis between the two substrates is of at least $32a$, depending on the liquid density. We considered system sizes with lateral cross sections raging from $25a$ to $80a$, meaning approximately $9\cdot 10^4$ to $9\cdot 10^5$ particles.

We perform isochoric Langevin dynamics simulations with a time step $\Delta t=0.01 \sqrt{m\sigma^{2}/\epsilon}$ and friction coefficient $\gamma=0.01 \Delta t^{-1}$ and track the progression of crystallization using the averaged local bond order parameters $\bar{q}_{4},\bar{q}_{6}$ proposed in \cite{Lechner:2008vz}. Their definition requires the computation of the complex vector $q_{l}(i)$
\begin{equation}
q_{lm}(i)=\frac{1}{N_{b}(i)}\sum_{j=1}^{N_{b}(i)}Y_{lm}(\bm{r}_{ij}) \;, 
\end{equation}
where $N_{b}(i)$ corresponds to the number of nearest neighbors of particle $i$ and $Y_{lm}(\bm{r}_{ij})$ reads as the spherical harmonics. Averaging over the neighbors of particle $i$ and particle $i$ itself
\begin{equation}
\bar{q}_{lm}(i)=\frac{1}{\tilde{N}_{b}(i)}\sum_{k=0}^{\tilde{N}_{b}(i)}q_{lm}(k),
\end{equation}
and summing over all the harmonics we finally get
\begin{equation}
\bar{q}_{l}(i)=\sqrt{\frac{4\pi}{2l+1}\sum_{m=-l}^{l}|\bar{q}_{lm}(i)|^{2}} \;.
\end{equation}
\end{appendix}

\bibliography{biblio,manual}{}

\begin{thebibliography}{10}
\bibitem{q6q6}%
  \BibitemOpen
  \bibinfo {journal} {A bond with a neighboring particle is named ``solid'' if
  the Steinheradt's order parameter product
  $\vec{q}_{6}\cdot\vec{q}_{6}>0.5$.}%
  \bibAnnoteFile{Stop}{q6q6}%
\end{thebibliography}

\end{document}